\documentclass[preprint2]{emulateapj}

\usepackage{appendix}
\usepackage{graphicx}
\usepackage{stmaryrd}
\usepackage{natbib}

\begin{document}

\newcommand{\vdag}{(v)^\dagger}
\newcommand{\1}{$^{1}$}
\newcommand{\2}{$^{2}$}
\newcommand{\3}{$^{3}$}
\newcommand{\4}{$^{4}$}
\newcommand{\5}{$^{5}$}
\newcommand{\6}{$^{6}$}
\newcommand{\7}{$^{7}$}
\newcommand{\8}{$^{8}$}
\newcommand{\9}{$^{9}$}
\newcommand{\0}{$^{10}$}
\newcommand{\myemail}{maselli@ifc.inaf.it}
\newcommand{\wse}{WISE}
\newcommand{\sax}{{\it Beppo}-SAX}
\newcommand{\cha}{{\it Chandra}}
\newcommand{\igr}{{\it INTEGRAL}}
\newcommand{\frm}{{\it Fermi}}
\newcommand{\sw}{{\it Swift}}
\newcommand{\xmm}{XMM-{\it Newton}}
\newcommand{\bz}{{\it Roma}-BZCAT~}
\newcommand{\eg}{{\em e.g.}}
\newcommand{\ie}{{\em i.e.}}
\newcommand{\nh}{$N_{\rm{H}}$}
\newcommand{\flux}{erg cm$^{-2}$ s$^{-1}$}
\newcommand{\sflux}{erg cm$^{-2}$ s$^{-1}$ Hz$^{-1}$}
\newcommand{\den}{$\times\,10^{\,20}$cm$^{-2}$}
\newcommand{\bi}{\textsc{BatImager}~}
\newcommand{\ra}{\textsc{ra}}
\newcommand{\dec}{\textsc{dec}}
\newcommand{\md}{$\times$ 10$^{-2}$}
\newcommand{\mt}{$\times$ 10$^{-3}$}
\newcommand{\st}{$^{\star}$}
\newcommand{\C}{$\circledcirc$}
\newcommand{\D}{$\boxcircle$}
\newcommand{\CC}{$\bigcirc$}
\newcommand{\B}{$\bullet$}
\newcommand{\A}{$\circ$}

\slugcomment{version \today: am}

\shorttitle{Searching for new $\gamma$-ray blazar candidates in the 3PBC with \wse}
\shortauthors{A. Maselli et al. 2013}

\title{Searching for new $\gamma$-ray blazar candidates \\ in the 3$^{rd}$ Palermo BAT Hard X-ray Catalog with \wse}

\author{A.~Maselli\altaffilmark{1}, F.~Massaro\altaffilmark{2},
  G.~Cusumano\altaffilmark{1}, R.~D'Abrusco\altaffilmark{3},
  V.~La~Parola\altaffilmark{1}, A.~Paggi\altaffilmark{3},
  A.~Segreto\altaffilmark{1}, Howard~A.~Smith\altaffilmark{3},
  G.~Tosti\altaffilmark{4,5}.}

\altaffiltext{1}{INAF-IASF Palermo, via U.~La~Malfa 153, I-90146 Palermo, Italy}
\altaffiltext{2}{SLAC National Accelerator Laboratory and Kavli Institute for Particle Astrophysics and Cosmology, 2575 Sand Hill Road, Menlo Park, CA 94025, USA}
\altaffiltext{3}{Harvard-Smithsonian Astrophysical Observatory, 60 Garden Street, Cambridge, MA 02138, USA}
\altaffiltext{4}{Dipartimento di Fisica, Universit\`a degli Studi di Perugia, I-06123 Perugia, Italy}
\altaffiltext{5}{Istituto Nazionale di Fisica Nucleare, Sezione di Perugia, I-06123 Perugia, Italy}

\begin{abstract}

We searched for $\gamma$-ray blazar candidates among the 382
unidentified hard X-ray sources of the 3$^{rd}$ Palermo BAT Catalog
(3PBC) obtained from the analysis of 66 months of \sw-BAT survey data
and listing 1586 sources.
We adopted a recently developed association method based on the
peculiar infrared colors which characterize the $\gamma$-ray blazars
included in the second catalog of active galactic nuclei detected by
the \frm\ Large Area Telescope (2LAC).
We used this method exploiting the data of the all-sky survey
performed by the Wide-Field Infrared Survey Explorer (\wse) to
establish correspondences between unidentified 3PBC sources and
\wse\ $\gamma$-ray blazar candidates located within the BAT positional
uncertainty region at 99\% confidence level.
We obtained a preliminary list of candidates for which we analysed all
the available data in the \sw~archive to complement the information in
the literature and in the radio, infrared and optical catalogs with
the information on their optical-UV and soft X-ray emission.
Requiring the presence of radio and soft X-ray counterparts consistent
with the infrared positions of the selected \wse\ sources, as well as a
blazar-like radio morphology, we finally obtained a list of 24
$\gamma$-ray blazar candidates.
\end{abstract}


\section{Introduction}
\label{sect:01}

One of the biggest challenges of the X-ray astronomy is understanding
the origin of the cosmic X-ray background
\cite{giacconi1962,mushotzky2000}.
A key issue to address this still open question is the association of
unidentified X-ray sources with their low- or high-energy counterparts,
which is a crucial step towards their identification and
classification.
The Burst Alert Telescope (BAT, \citealp{barthelmy2005}) on board the
\sw~observatory \cite{gehrels2004}, thanks to its wide field of view
($\sim$ 1.4~sr) coupled with a large collecting area
($\sim$~5200~cm$^2$), is performing since its launch an all-sky survey
in the 15--150~keV energy range.
This survey provides an unprecedented view of hard X-ray detected
active galactic nuclei (AGNs) (\eg, \citealp{burlon2011}),
corresponding to a significant fraction (\ie, $\sim$ 60\%,
\citealp{ajello2012}) of the BAT detected sources, so increasing our
knowledge on their contribution to the cosmic X-ray background (\ie,
\citealp{gilli2007,ajello2009}).

The 3$^{rd}$ Palermo BAT Hard X-ray
Catalog\footnote{http://bat.ifc.inaf.it}, obtained from the analysis
of 66 months of BAT all-sky data with the {\sc batimager} software
\cite{segreto2010} and covering 50\% of the sky down to a 15--150 keV
flux limit of $7.4 \cdot 10^{-12}$ \flux, lists 1586 sources.
In the past years, thanks to follow-up observations performed by the
\sw~mission itself with its narrow field instruments UVOT
\cite{roming2005} and XRT \cite{burrows2005}, a lot of previously
unidentified BAT sources have been associated with a counterpart;
however, 382 out of 1586 ($\sim$~24\%) of them are still unidentified.
For comparison, the 4$^{th}$
\igr\ catalog\footnote{http://irfu.cea.fr/Sap/IGR-Sources/}
\cite{bird2010} of the all-sky survey performed with the IBIS/ISGRI
\cite{ubertini2003,lebrun2003} instrument lists 723 sources detected
in the 17-100~keV energy range, $\sim$~30\% of which are unidentified.
The largest fraction of the associated \igr\ sources (\ie,
$\sim$~35\%) are AGNs, with respect to $\sim$~31\% of identified
Galactic sources.
The fraction of unidentified sources is probably larger in
\igr~IBIS/ISGRI compared to \sw-BAT because the former spends more
time on the Galactic plane, where source identification is naturally
more challenging.

A limited fraction ($\sim$~15\%) of the AGNs detected by BAT and
included in the 3PBC is constituted by blazars which represent, in
turn, the largest known population of $\gamma$-ray sources detected by
the \frm\ Large Area Telescope (LAT, \citealp{atwood2009}).
Blazars are peculiar AGNs whose emission is dominated by non-thermal
radiation over the entire electromagnetic spectrum.
They belong to the radio-loud class of AGNs and are characterized by
flat radio spectra, peculiar infrared (IR) colors
\cite{massarof2011a}, apparent superluminal motion, and a typical
double humped spectral energy distribution (SED)
(\eg, \citealp{urry1995}).
In 1978 Blandford \& Rees proposed to interpret the blazar emission as
arising from particles accelerated in relativistic jets, observed at a
small angle with respect to the line of sight.
Thus, their SED has been described in terms of synchrotron radiation
from radio through the soft X-rays, and of inverse Compton scattering
of low-energy photons from the hard X-rays to the $\gamma$-ray band
\cite{marscher1985,dermer2002}.
Blazars are divided in two subclasses: the BL~Lac objects (BL~Lacs)
characterized by weak or absent optical emission lines, and the flat
spectrum radio quasars (FSRQs) showing broad emission lines in their
optical spectrum (\eg, \citealp{stickel1991,stocke1991}).
The \bz Multi-Frequency Catalog of
Blazars\footnote{http://www.asdc.asi.it/bzcat/}
\cite{massaroe2009,massaroe2011b} is, at present, the most complete
list of blazars including certified BL Lacs and FSRQs as well as
blazars with uncertain classification.

Using the recent all-sky survey performed by the Wide-Field Infrared
Survey Explorer (\wse, \citealp{wright2010}) D'Abrusco et~al.~(2012a)
showed that $\gamma$-ray emitting blazars detected by \frm-LAT and
included in the second LAT AGN catalog (2LAC, \citealp
{ackermann2011}) can be well separated from other Galactic and
extragalactic sources on the basis of their IR colors.
This analysis confirmed and strengthened the results obtained
previously by Massaro et~al.~(2011a) on the basis of the
\wse\ preliminary data release and constituted the first step to
develop a new association procedure for the unidentified $\gamma$-ray
sources \cite{massarof2012a}.
Applying this association method, which is able to recognize if there
is a $\gamma$-ray blazar candidate within their \frm\ location
uncertainty, Massaro et~al.~(2012b) assigned a $\gamma$-ray blazar
candidate counterpart to 156 out of 313 unidentified $\gamma$-ray
sources analysed, having the same IR properties of the $\gamma$-ray
emitting blazars.
Several methods have been used in the past to identify blazars in
radio and X-ray surveys as, for example, those proposed by Perlman
et~al.~(1998) for the Deep X-Ray Radio Blazar Survey (DXRBS) or by
Laurent-Muehleisen et~al.~(1999) in the case of the RGB sample as well
as the more recent case of Beckmann et~al.~(2003) for the HRX-BL Lac
sample.
In these cases, the combination of the radio-to-optical and/or the
optical-to-X-ray spectral indices was a good indicator of blazar-like
sources (see also \citealp{giommi1999,turriziani2007}).
However, the new approach based on the IR peculiar colors of blazars
(\citealp{massarof2012a}) can be applied to the whole sky when the
lack of multifrequency information, mostly in the radio and in the
optical bands, does not allow to adopt the above criteria.

The main scientific objective of the present paper is the search for
\wse\ $\gamma$-ray blazar candidates that could be possible
counterparts of unidentified BAT objects listed in 3$^{rd}$ Palermo
BAT Catalog.
At this purpose we applied the association method developed by Massaro
et~al.~(2012a): then, for the established 3PBC-\wse\ correspondences
we searched for \sw~pointed observations to reveal the presence of an
optical--UV and/or soft X-ray counterpart to each \wse\ $\gamma$-ray
blazar candidate that are crucial to address a correct classification.
A similar investigation has been carried out \cite{massarof2012c}
searching among the unidentified sources of the 4$^{th}$
\igr\ catalog, leading to the definition of 18 $\gamma$-ray blazar
candidates.

This paper is organized as follows: the application of the
\wse\ association method to blazars detected by both \sw-BAT and
\frm-LAT, as well as to the unidentified 3PBC~objects, is illustrated in
Section~\ref{sect:02}; the multifrequency properties of the obtained
\wse\ candidates, including details about the \sw~data reduction for
both the UVOT and XRT telescopes, are described in
Section~\ref{sect:03}.
The final list of the 3PBC-\wse~correspondences, obtained after
establishing more restrictive selection criteria based on the peculiar
emission of blazars, is presented in Section~\ref{sect:04} together
with additional comments and details on some selected candidates.
Finally, Section~\ref{sect:05} is devoted to our summary and
conclusions.

The nominal \wse\ bands are [3.4], [4.6], [12], and [22] $\mu$m;
\wse\ colors are [3.4]-[4.6] mag, [4.6]-[12] mag, and [12]-[22] mag.
Unless stated otherwise, we use c.g.s. units throughout.

\section{The \wse\ Gamma-ray Strip association procedure and the 3PBC sources}
\label{sect:02}

In this Section we describe the application of the \wse\ association
procedure to some, opportunely selected, samples of BAT sources
included in the 3PBC.
Some basic details of this procedure can be found in the
Appendix; a complete description is given in Massaro
et~al.~(2012a,b) and additional details are also illustrated in
D'Abrusco et~al.~(2013).

The method is based on the parameterization of the {\it Gamma-ray
  Strip}, the very narrow region of the IR color-color space in
which the \wse\ counterparts, detected in all of the four \wse\ bands,
of the $\gamma$-ray blazars included in the 2LAC \cite{ackermann2011}
are located.
The position of a generic \wse\ source in the IR color-color space
with respect to the strip is represented by a parameter $s$,
normalized to vary in the range between 0 (\wse\ source definitely
outside the strip) and 1.
Three classes (A, B, and C) have been opportunely defined on the basis
of the values of the $s$ parameter: high values are typical of class~A
while low values correspond to class~C.
For each 3PBC source a circular searching region centered at its
position as provided in the 3PBC and with a radius
equal to the positional uncertainty at 99\% confidence level is
defined.
This uncertainty varies in the range 0.8 -- 8.4 arcmin according to
the S/N ratio of the source, with a mean value of (5.2~$\pm$~1.4)
arcmin.
Then, all the \wse\ sources within this searching region are ranked
according to their $s$ parameter: of the obtained list of candidates
we indicate, as a rule, our best choice as the \wse\ source of higher
class with the smallest angular separation from the BAT position.

\begin{figure}[hbtp]
\begin{center}
\includegraphics[width=8.5cm]{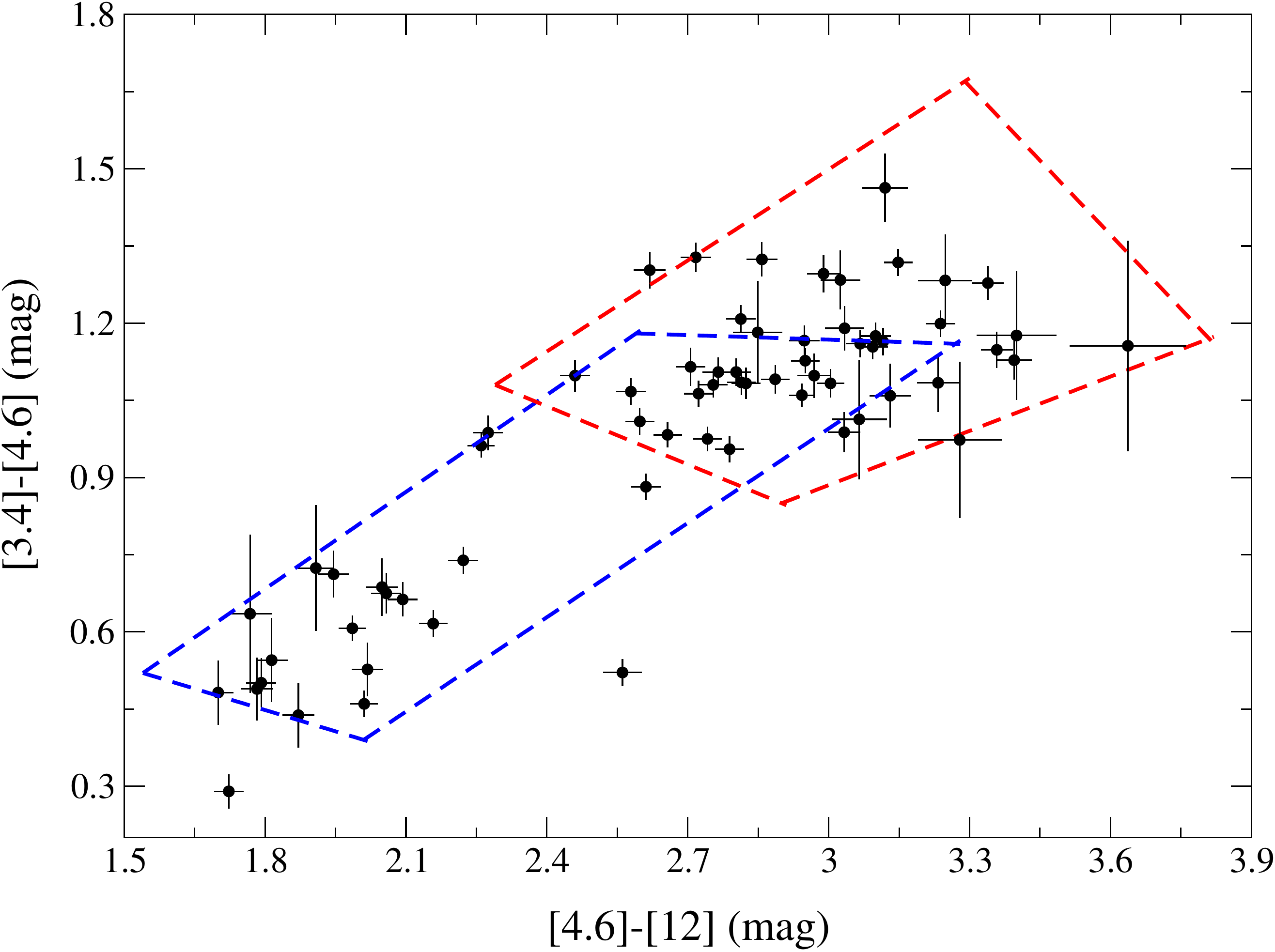}
\end{center}
\caption{[3.4]-[4.6]-[12]~$\mu$m color-color diagram showing the
  location of the \wse\ counterparts of the BAT-LAT blazars with
  respect to the boundaries of the \wse\ Gamma-ray Strip. The BL~Lac
  objects and FSRQs subregions of the \wse\ Gamma-ray Strip are marked
  by blue and red dashed lines, respectively.}
\label{fig:01}
\end{figure}

First, we report the results of the application of the above
described procedure to the \bz certified blazars detected by both
\sw-BAT and \frm-LAT.
We computed how many of these blazars are found within the boundaries
of the Gamma-ray Strip, as expected: in this way we could evaluate the
reliability of the association method.
The number of \bz certified blazars that are present in the 3PBC is
125, which corresponds to a modest fraction ($\sim$~10\%) of the 1204
identified 3PBC objects.
According to the 2LAC \cite{ackermann2011} 67 out of these 125 are
associated to a \frm-LAT source.
We treated these 67 BAT-LAT blazars as if they were unidentified and
applied our procedure: in particular, we verified that the
positionally closest source belonging to the highest class was the
same associated in the 3PBC.
Their location with respect to the \wse\ Gamma-ray Strip is shown in
Figure~\ref{fig:01}: for this particular 2-dimensional projection of
the \wse\ color-color space only two sources are completely outside
the \wse\ Gamma-ray Strip ($s_{12}=0$).
Considering also the remaining two 2D-projections, the method was able
to confirm 62 out of 67 BAT-LAT blazars: this implies a high
capability ($\sim$~93\%) of validating $\gamma$-ray blazars through
our association procedure.
These 62 blazars are distinguished in 17 sources of class~A
($\simeq$~28\%), 20 of class~B ($\simeq$~32\%), and 25 of class~C
($\simeq$~40\%).

Then, we applied our association procedure to the unidentified sources
included in the 3PBC.
From a general point of view, hard X-ray sources like those
constituting the 3PBC can be considered identified when a counterpart
at some other energy band (radio, IR, soft X rays), whose coordinates
are provided with arcsecond precision, has been associated to them and
when this counterpart has been opportunely classified so that the
nature of these sources is well established.
Therefore, the 382 3PBC unidentified objects can be further distinguished in 
two groups: 222 unassociated sources, for which a counterpart has not
been found, and 160 unclassified objects that are generic radio, soft
X-ray and $\gamma$-ray sources for which a lower energy counterpart
has been assigned in the 3PBC but no classification has been yet
provided.
Applying our association procedure to all the 382 unidentified objects
we obtained a \wse~$\gamma$-ray blazar candidate for 22 unassociated
and 39 unclassified 3PBC sources.

\begin{figure*}[] 
\begin{center}
\includegraphics[height=6cm]{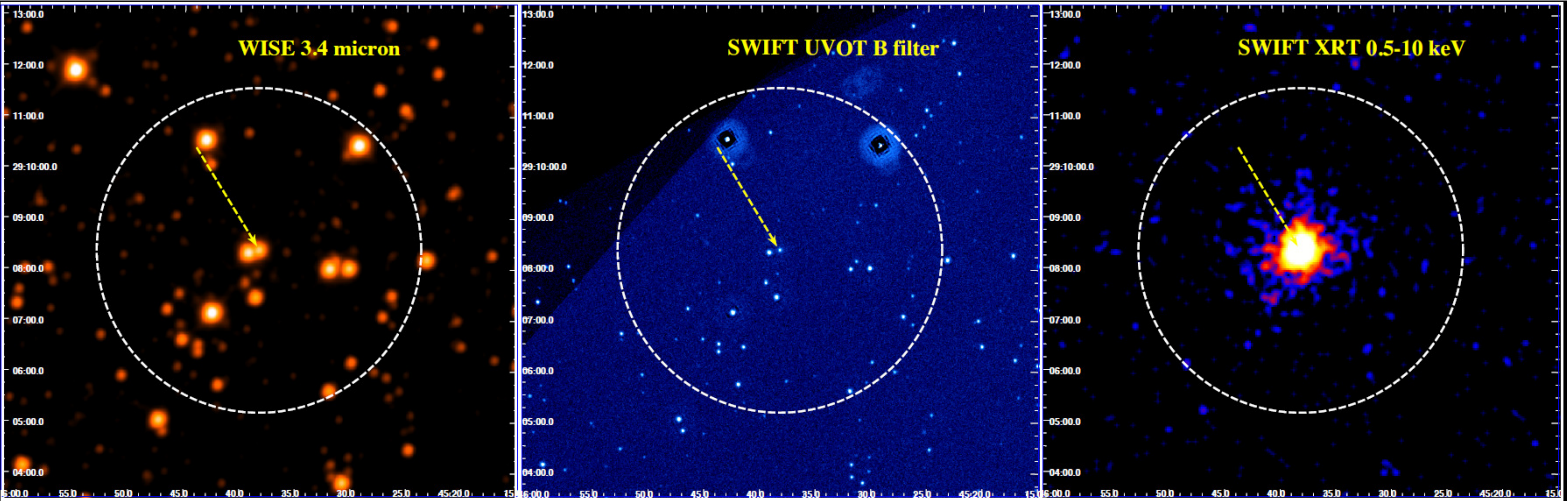}
\caption{The $\gamma$-ray blazar candidate \wse\ J174538.26$+$290822.2
  in the infrared, optical and soft X-ray maps. The dashed circle in
  the soft X-ray map corresponds to the BAT positional uncertainty
  region at 99\% confidence level.}
\label{fig:02}
\end{center}
\end{figure*}

\section{Multifrequency properties \\ of the \wse\ $\gamma$-ray blazar candidates}
\label{sect:03}

Once that a \wse\ $\gamma$-ray blazar candidate was found for a 3PBC
source by the application of our method, we investigated its possible
emission in energy ranges different from the infrared and the hard
X-ray ones by searching for the available information in the
literature and for observations in the \sw~database to cover the
optical-UV as well as the soft X-ray bands.

The values of the radio flux density $F_r$ were extracted from the
main radio surveys such as FIRST (\citealp{white1997}), NVSS
(\citealp{condon1998}), PMN (\citealp{wright1994}) and SUMMS
(\citealp{mauch2003}).
We searched for radio sources included in a circle centered at the
position of the \wse\ candidate and with a conservative radius of
30$\arcsec$ to take into account the possibly low sensitivity of some
surveys as well as the intrinsical weakness of some radio sources.
When available, we also checked the radio morphology of the source to
discard radio galaxies with an evident FRI/FRII morphology
\cite{fanaroff1974} and focus on sources with a compact core or with a
core/single-sided jet structure characteristic of blazars.

Moreover, we searched for candidates in the field of the Sloan Digital
Sky Survey-Data~Release~9 (SDSS-DR9, \citealp{paris2012}) to obtain
the values of the photometry in the $u$, $g$, $r$, $i$ and $z$ filters.
As recently shown by Massaro et~al.~(2012d), the $u-r$ color index is
a simple and well-suited parameter to measure the AGN contribution to
the source emission with respect to the host galaxy. 
Therefore, it can be used to introduce a new classification scheme for
blazars distinguishing between galaxy-dominated blazars ($u-r >
1.4$~mag) and nuclear-dominated blazars ($u-r <1.4$~mag),
respectively.
We calculated the intrinsic $u-r$ values adopting their formula:
\begin{equation}
\label{ur}
(u-r) = (u-r)_{obs} -0.81 A_r
\end{equation}
where $A_r$ is the extinction in the $r$~band provided in the SDSS database.

\subsection{Swift UVOT and XRT observations}
\label{sect:03:1}

We searched in the \sw~archive at the position of each
\wse~$\gamma$-ray blazar candidate within a circular region with a
radius of 15\arcmin, wide enough to take into account the eventuality
that the source of interest is not necessarily close to the center of
the field of the \sw~observation.

We carried out the photometry of UVOT data adopting the following
procedure: we summed the available frames with the {\sc uvotimsum}
task and obtained a single image in FITS format in the corresponding
filter.
We defined for each source a circular region with the recommended
radius of 5\arcsec\ adopted for the calibration of counts to
magnitudes \cite{poole2008}.
For the background region a much larger value of the radius, typically
20\arcsec, was considered.
Then, we used the task {\sc uvotsource} adopting a 3~$\sigma$ level of
significance to compute the background limit, and obtained the values
of the photometry in the Vega System; we took into account both
statistic and systematic errors.

We analysed X-ray data with the aim of determining with the best
possible accuracy the position of the X-ray source.
In the case of several observations for the same source we downloaded
the data corresponding to the longer XRT exposure time.
However, when a longer exposure time was needed to establish the
position of the X-ray source with sufficient accuracy due to a modest
count rate, we downloaded all the available observations and summed
the corresponding event files to obtain a single frame.
We reduced XRT data with the XRT Data Analysis Software ({\sc
  XRTDAS}~v2.8.0) included and distributed within the HEASoft~v6.12
package.
Cleaned event files, as well as the other analysis products, were
obtained with the {\sc xrtpipeline} task.
Multiple event files from different observations were summed with the
{\sc xselect} task.
Then, the event files were examined with {\sc ximage} and a
preliminary source detection was carried out with a sliding-cell
method to reveal significant X-ray sources in the field.
The coordinates of the eventual X-ray source closest to the
\wse~source were instead determined using the {\sc xrtcentroid} task,
providing in this way an error radius to the X-ray position.
We found that the position of XRT sources is known with a precision of
a few arcseconds, ranging from 3.5\arcsec~up to 8\arcsec, depending on
the count rate of the source and on the exposure time of the
observation.

We used the images acquired in the UVOT filters to verify the
consistency of the \wse\ and the XRT source with each other and with
the optical-UV counterpart, when detected.
We examined the images with DS9 and defined the region files to locate
the position, with their errors, of both \wse~sources and relevant,
previously detected, XRT sources in the field.
For the position of the \wse~sources we adopted, for the sake of
simplicity, a conservative error radius of 2\arcsec (see
\citealp{cutri2012} for further details).

\section{Results and source details}
\label{sect:04}

\subsection{More stringent criteria \\ on the selection of blazar candidates}
\label{sect:04:1}

Due to the presence of multifrequency observations for several
unidentified BAT sources we established more restrictive criteria
with respect to those adopted for the \frm\ unidentified $\gamma$-ray
sources \cite{massarof2012b} to select our list of $\gamma$-ray
blazar candidates.

\begin{figure}[hbtp]
\begin{center}
\includegraphics[width=8cm,angle=-90]{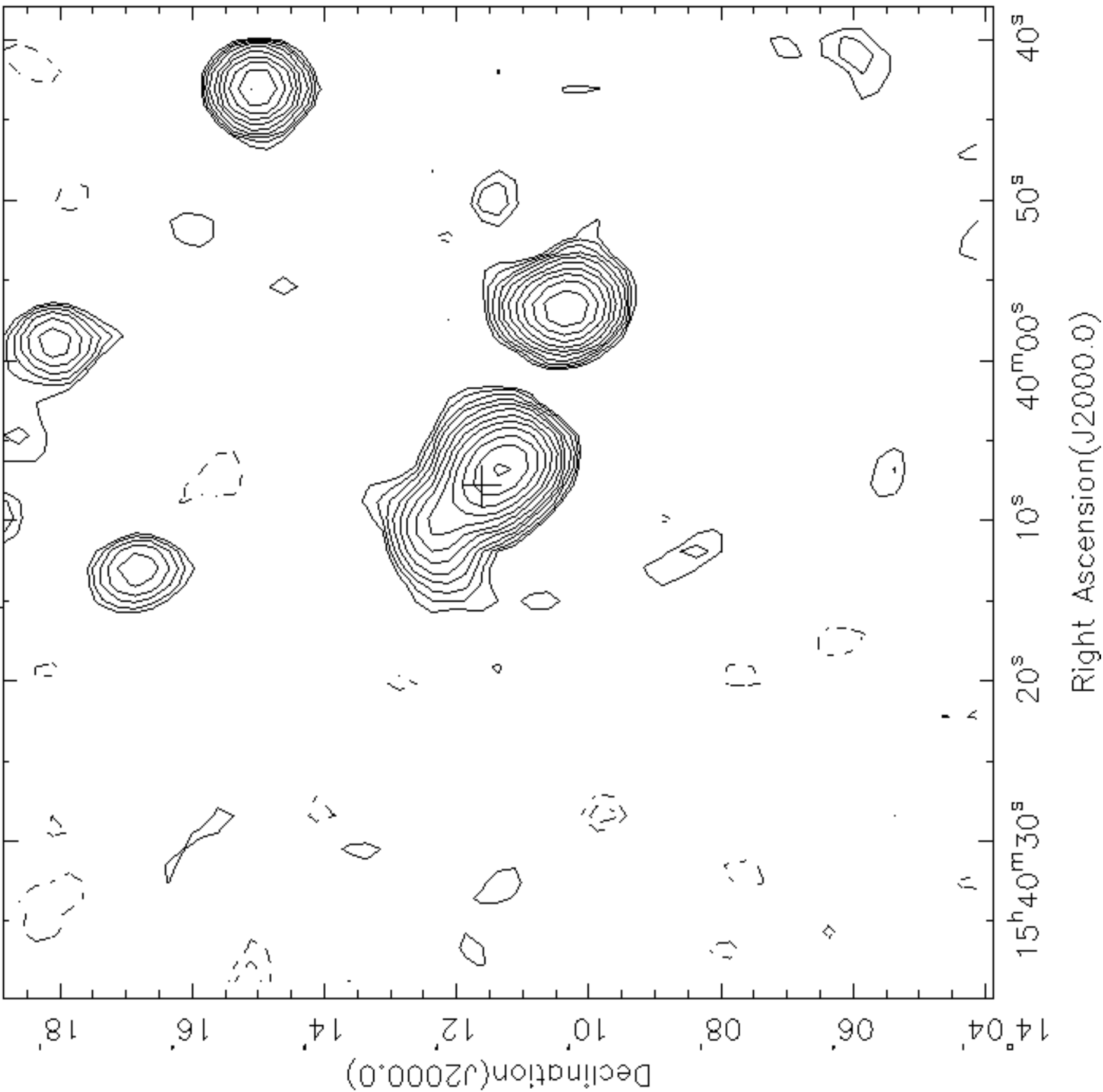}
\end{center}
\caption{The NVSS map showing the core/single-sided jet structure of
  the radio counterpart of \wse\ J154007.85$+$141137.2.}
\label{fig:03}
\end{figure}

First, we required that the position of the \wse\ $\gamma$-ray blazar
candidates was included in the uncertainty region corresponding to the
soft X-ray emission by \sw-XRT or by other X-ray telescopes.
Second, in addition to peculiar IR colors as those of the $\gamma$-ray
blazars, we required the $\gamma$-ray blazar candidates to have a
radio counterpart.
Moreover, we checked the radio morphology of the source in the NVSS
map, when available, and confirmed candidates with a compact core or
with a core/single-sided jet structure as shown in
Figure~\ref{fig:03}.
In the case of \wse\ J010930.21$+$085748.1 we found a multicomponent
structure compatible with the signature of the jet on a kpc scale;
such occurrence has been already evidenced for other well studied
sources like 3C~273 or 3C~454.3, where the jet has been imaged also in
the X rays \cite{massarof2011b} with the \cha~ X-ray satellite.

\begin{table}
\scriptsize
\caption{The correspondences between unassociated 3PBC sources
  and \wse\ $\gamma$-ray blazar candidates with a core-dominated NVSS
  radio counterpart but no soft X-ray detection. 1) The 3PBC name, 2)
  the \wse\ name, 3) the class of the candidate, 4) its distance
  $d^{\star}$ from the 3PBC source and 5) its radio flux density
  $F_r$.}
\label{table:1}
\begin{center}
\begin{tabular}{ccccc}
\hline
~\\
3PBC name      &        WISE name      & class & $d^{\star}$ & $F_r$ \\ 
               &                       &       &  (arcsec)  & (mJy) \\
~\\                                                                 
\hline                                                              
~\\                                                                 
J0744.7$-$2348 & J074450.96$-$235014.8 &   C   &   110.6    &  3.0 \\
J1922.9$+$2648 & J192300.98$+$265054.8 &   C   &   148.5    & 45.5 \\
J2131.3$+$6100 & J213130.14$+$605752.2 &   A   &   202.9    & 11.7 \\
~\\
\hline
\end{tabular}
\end{center}
\end{table}

The adoption of these criteria severely reduced the number of
candidates with respect to our preliminary list, particularly among
those provided for unassociated 3PBC sources.
In fact, the largest part of candidates that fulfilled them correspond
to unclassified objects, with only a single candidate given for an
unassociated 3PBC source.
This is mainly due to the large difference between unassociated and
unclassified objects in the relative fraction of objects for which at
least a \sw~observation is available.
This gap is, in turn, the natural consequence of the fact that a large
fraction of the counterparts to the hard X-ray sources in the 3PBC has
been found thanks to dedicated \sw~observational campaigns.

We report in Table~\ref{table:1} three candidates included in our
preliminary list with a core-dominated radio counterpart in the NVSS
but, unfortunately, no available detection in the soft X-ray band,
neither by \sw\ nor by other X-ray satellites.
The observation of this small number of objects by \sw\ or by other
soft X-ray telescopes would be a very simple and helpful step for
addressing their correct classification.

\subsection{The list of $\gamma$-ray blazar candidates}
\label{sect:04:2}

From the application of our association procedure and the further
adoption of our selection criteria we obtained 24 $\gamma$-ray blazar
candidates.
The list of 3PBC-\wse~correspondences, sorted following the right
ascension of the 3PBC sources, is presented in Table~\ref{table:2}.
For each correspondence it is reported 1) the name of the 3PBC source
and 2) one of the names with which the 3PBC~counterpart is addressed
in the main astronomical databases
(NED\footnote{http://ned.ipac.caltech.edu/},
SIMBAD\footnote{http://simbad.u-strasbg.fr/simbad/}); 3) the name of
the \wse\ $\gamma$-ray blazar candidate, 4) the class designated by
our association procedure and 5) its angular separation $d$ from the
3PBC counterpart.
The redshift of the \wse~source as found in the literature with the
corresponding reference is reported in column~6; the origin of the
redshift is spectroscopic in all the cases with the exception of
\wse\ J174201.50$-$605512.1 \cite{burgess2006} for which it is
photometric.
The values of the radio flux density $F_r$ from the main radio surveys
are reported in column~7.
The distance between the coordinates of the \wse\ candidate and the
radio source is tipically lower than 3$\arcsec$: for three
\wse\ candidates (J044047.72$+$273947.1, J192630.21$+$413305.0 and
J221409.17$-$255749.1) we found a higher value, in any case lower than
15$\arcsec$, but we considered the possibility that this discrepancy
is due to the low flux of the radio source, of the order of a few mJy.
The intrinsic $u-r$ color index, calculated adopting Eq.~(\ref{ur}), is
available for eight of these candidates and is reported in column~8.
Six of these values are lower than 0.9~mag, configuring the sources
as nuclear-dominated blazar candidates (see Section~\ref{sect:03}).
In column~9 it is summarized the comparison between the position of the
\wse~source, represented by a small inner circle, and the position of
the closest X-ray source in the field of the available XRT
observations, represented by a circle with a larger radius.
Two concentric circles indicate that the position of the \wse~source
with its error is completely included in the uncertainty region of the
XRT source. 
This agreement occurs for all the selected candidates, including the
three sources with no \sw\ observation but included in the ROSAT
All-Sky Survey Bright Source Catalog (1RXS) \cite{voges1999}.
In these cases the larger circle has been replaced by a square in the
symbols adopted in column~9.

\begin{figure}[] 
\begin{center}
\includegraphics[width=8.5cm]{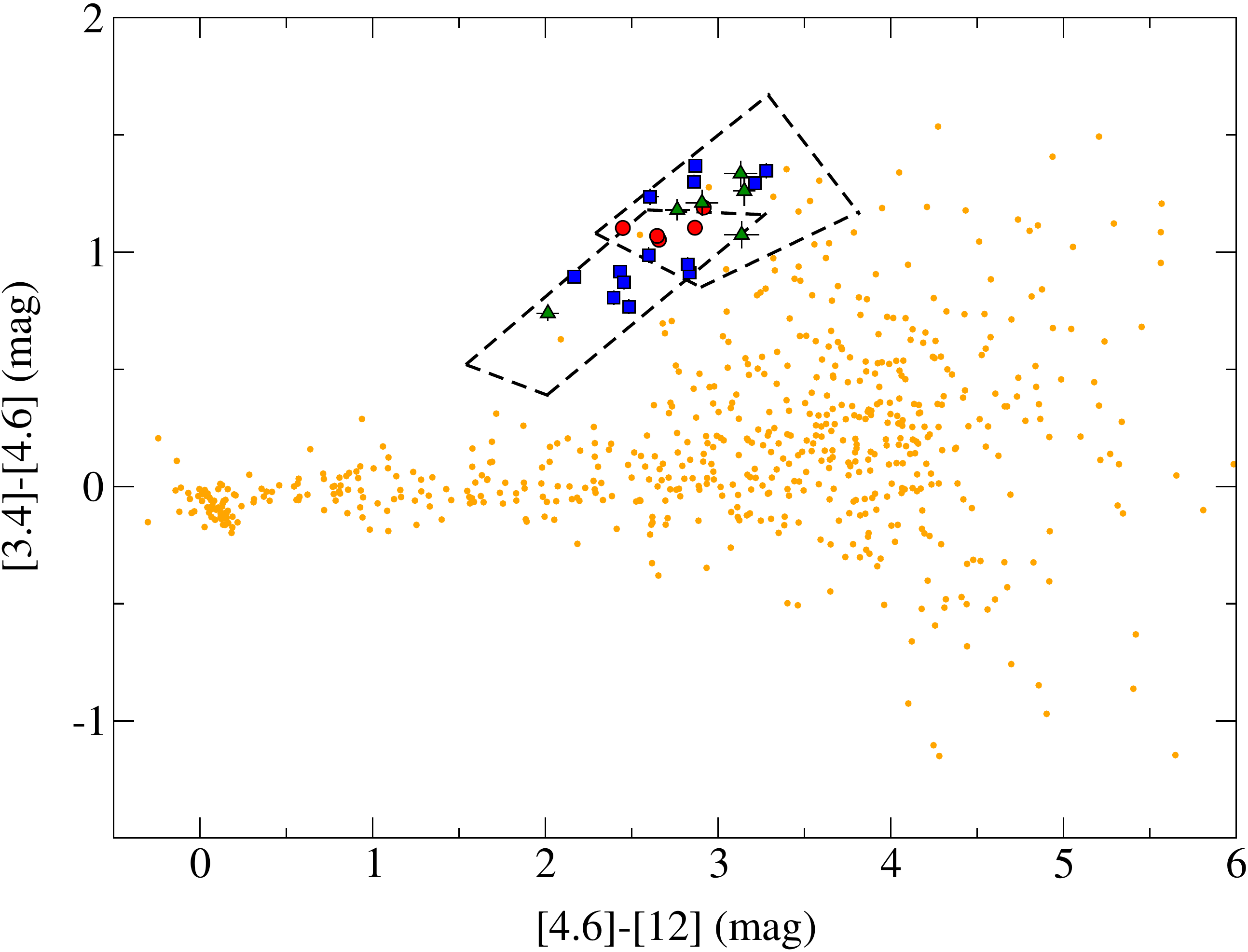}
\caption{The [3.4]-[4.6]-[12] $\mu$m diagram showing the IR colors of
  the 24 $\gamma$-ray blazar candidates distinguished among class~A
  (red circles), class~B (blue squares), and class~C (green triangles)
  candidates, respectively. The boundaries corresponding to the
  projection of the \wse\ Gamma-ray strip in this color-color plane
  are also reported. Orange circles represent $\sim$~600 generic IR
  sources detected within the positional uncertainty regions of the
  unidentified 3PBC sources.}
\label{fig:04}
\end{center}
\end{figure}

The position of our 24 selected candidates in the [3.4]-[4.6]-[12]
$\mu$m diagram with respect to the boundaries corresponding to the
projection of the \wse\ Gamma-ray strip is plotted in
Fig.~\ref{fig:04}.
We also reported, for comparison, the IR colors of $\sim$~600 generic
IR sources detected within the positional uncertainty regions of the
unidentified 3PBC sources and that are not selected as $\gamma$-ray
blazar candidates by our method.
We note that a few sources have IR colors similar to those of
$\gamma$-ray blazars but are not selected by our method as, in the
other 2 color-color planes, they lie outside the boundaries
corresponding to the projections of the \wse\ Gamma-ray Strip or the
values of their strip parameters $s_{23}$ or $s_{34}$ are very low.

\begin{figure}[] 
\begin{center}
\includegraphics[height=6.5cm]{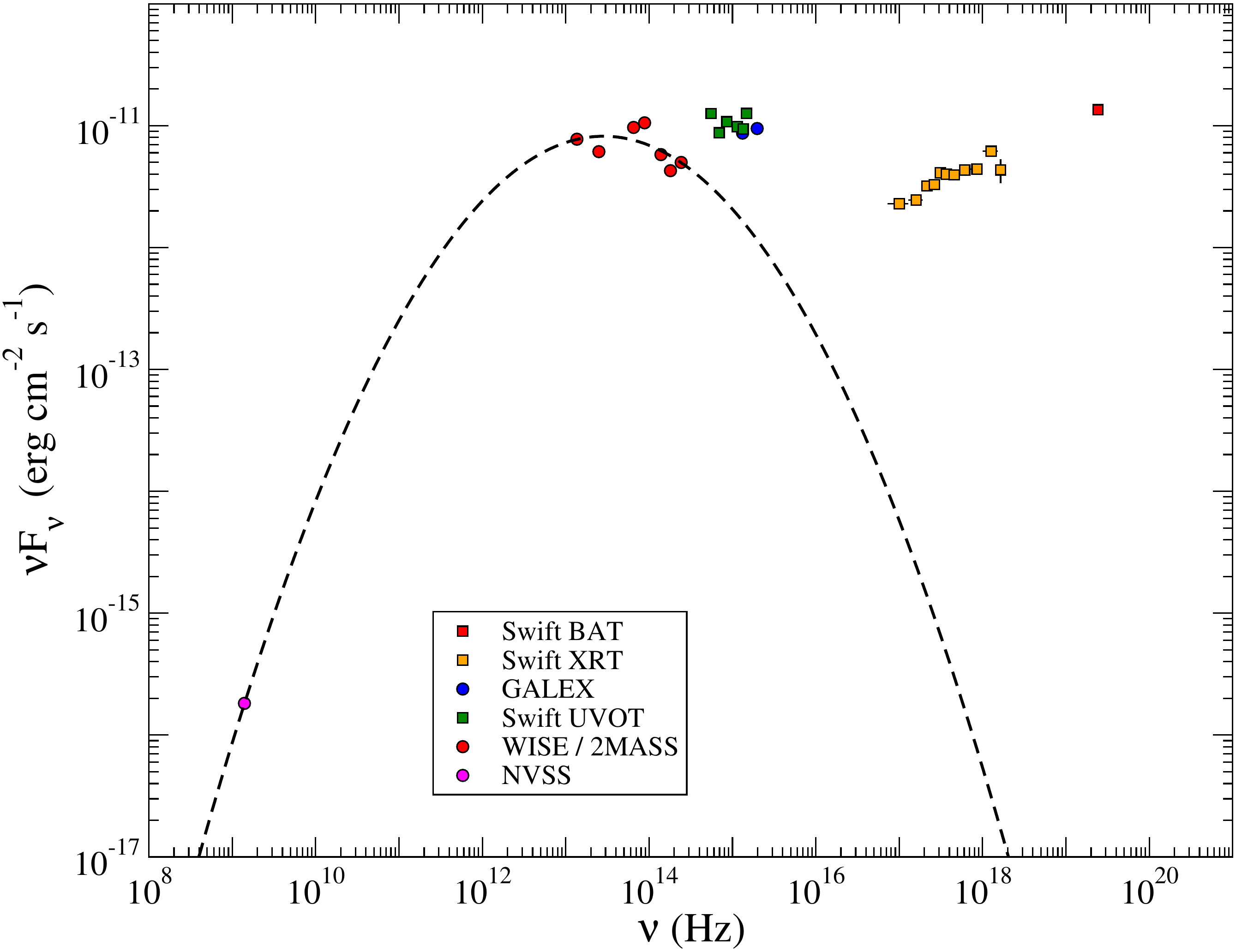}
\caption{The SED of the $\gamma$-ray blazar candidate
  \wse\ J174538.26$+$290822.2. Both the IR and the optical data
    have been de-reddened for Galactic absorption.}
\label{fig:05}
\end{center}
\end{figure}

The comparison of the maps in the infrared, in one of the UVOT
filters, and in the soft X-ray band for one of these candidates is
shown in Figure~\ref{fig:02}.
For the same source we also show the SED in Figure~\ref{fig:05}; a
log-parabolic model fits the radio and the IR data while the
data in the optical filters, which are above the fit and are not
simultaneous with the IR ones, can be explained as the signature
of the intrinsic source's variability.
The X-ray data in the soft (XRT) and in the hard (BAT) bands are
consistent with each other and can be interpreted as the rising branch
of the inverse Compton component.

We note among these candidates the presence of two sources
already included in the \frm\ catalogs.
\wse\ J013750.47$+$581411.3 has been detected in both the 1FGL
\cite{abdo2010b} and the 2FGL \cite{nolan2012}.
It has been already proposed, among others, by Maselli et al.~(2011)
as a blazar source and in particular as a high-energy synchrotron peak
(HSP, \citealp{abdo2010a}) blazar after the analysis of its SED.
\wse\ J150838.93$-$495302.2 has been instead addressed by Landi
et~al.~(2012) and by Tuerler et~al.~(2012) as the counterpart of the
2FGL source J1508.5$-$4957; there is no corresponding $\gamma$-ray
source in the 1FGL.
The analysis of the spectral index in different energy bands like the
radio \cite{massardi2008} the microwaves \cite{planck2011} and the
soft X-rays \cite{landi2012} configures this object as a
low-energy synchrotron peak (LSP, \citealp{abdo2010a}) source.
The fact that only two of our candidates have been already included in
the \frm\ catalogs is not unexpected and may be justified, in first
instance, considering the extreme variability of blazar sources.
Similar cases have been already shown by the two recently discovered
$\gamma$-ray sources \frm\ J1350$-$1140 \cite{torresi2011} and \frm\
J1717$-$5156 \cite{schinzel2012}, detected in a flaring state well
above the flux limit of the 2FGL on day timescale but not present in
the 2FGL itself; these sources have been associated using the \wse\ IR
colors as reported in Massaro et~al.~(2012e) and Paggi
et~al.~(2012a), respectively.

All the 21 candidates that have been observed by \sw, with only one
exception, have been detected in at least one of the optical-UV
\sw-UVOT filters: details about their photometry are given in
Table~\ref{table:3}.
Several objects have been found with a very low value of the Galactic
latitude: for this reason we preferred to report in the
Table~\ref{table:3} the value of the magnitude not corrected for
Galactic extinction together with the corresponding E(B-V) value as
derived by the Infra-Red Science
archive\footnote{http://irsa.ipac.caltech.edu/applications/DUST/}(IRSA).
The lack of a detection in any of the UVOT filters for
\wse\ J181250.94-064825.4 is most probably due to the combined effect
of its intrinsic weakness, the low value of its Galactic latitude, and
an insufficient exposure time of the observations.

\begin{figure}[thbp]
\begin{center}
\includegraphics[width=8.5cm]{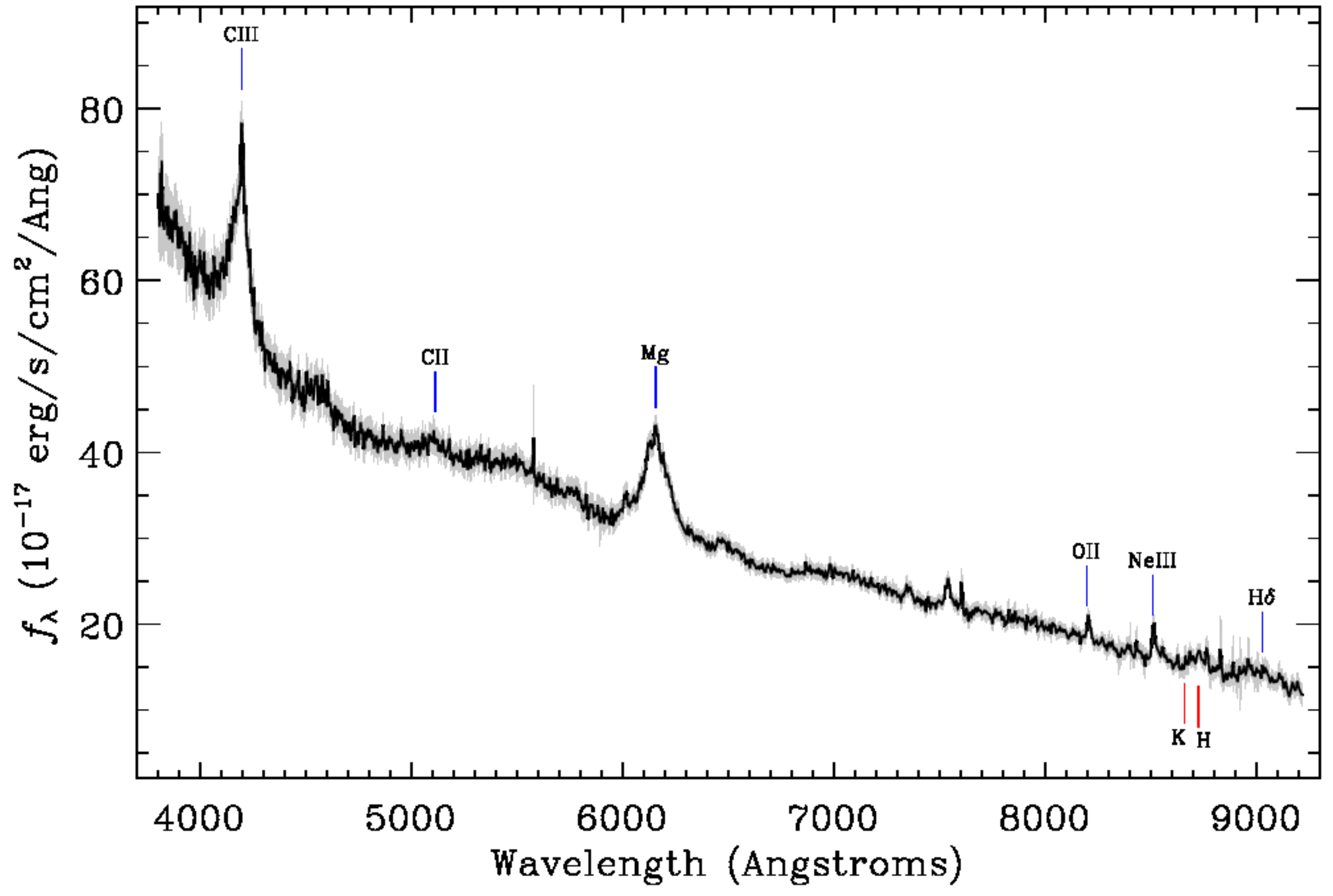}
\end{center}
\caption{The optical spectrum of \wse\ J124021.14$+$350259.0 provided
  by the SDSS-DR9.}
\label{fig:06}
\end{figure}

Among the candidates not observed by \sw\ we address
\wse\ J124021.14$+$350259.0 which benefits of a quite rich information
such as a considerable radio emission (230.8~mJy in the NVSS), a very
low value of its color index ($u-r = 0.20$~mag) and an optical
spectrum as shown in Figure~\ref{fig:06}.
With its large and strong emission lines and its continuum blueward
emission this candidate has indeed many requisites to be classified as
a flat spectrum radio quasar.

\subsection{Spurious associations}
\label{sect:04:3}

The number of generic IR sources lying within the positional
uncertainty region at 99\% level of confidence of each unidentified
3PBC source is definitely larger than 1, since the source density of
\wse\ sources is very large \citep{wright2010}.
However, to estimate the probability that our associations are
spurious we need to verify the chance to have an IR source not only of
class~A, B or C (therefore with oppurtune color indices) but also
with a radio counterpart in a random region of the sky.
Thus, we considered the following approach.

We created a list of 100 {\it fake hard X-ray} positions randomly
selecting 100 unidentified 3PBC sources that lie in the northern
hemisphere (\ie, Dec. $>$ 5 deg) and shifting their positions by 30$'$
in a random direction of the sky; the adoption of this threshold on
the Declination guarantees that NVSS radio data are potentially
available for all of them.
We assigned to each {\it fake} hard X-ray source an error radius
randomly chosen from the distribution of those corresponding to the
unidentified BAT sources.
We also verified that no 2FGL source was included within this error
radius.
Applying our association procedure to search for \wse\ $\gamma$-ray
blazar candidates of A, B, and C classes within the positional
uncertainty region of these 100 {\it fake} positions we found no
\wse\ source of class~A, 3 sources of class~B and 16 of class~C; all
these {\it fake} associations are unique in the sense that we never
found more than one \wse\ source either of class A, B or C for the
same {\it fake} position.

According to our restrictive criteria a \wse\ source can be considered
a valid $\gamma$-ray blazar candidate only if provided with a radio
counterpart.
Consequently, we searched for the NVSS counterpart of each of the
class~B and class~C sources corresponding to a {\it fake} position.
We found that only one source of class~B and only one of class~C have
an NVSS counterpart out of 100 {\it fake} hard X-ray sources.

We repeated the above described procedure 1000 times and we concluded
that the probability of chance coincidence is of the order of 1\%
for sources of class~B and class~C while it is lower than 0.1\% for
sources of class~A.
We emphasize that these probability estimates are only upper limits: a
detailed \frm\ analysis is indeed necessary to verify that no
significant $\gamma$-ray emission is due to the {\it fake} source of
class~B and to that of class~C with a radio counterpart obtained from the
application of our procedure to the list of {\it fake} positions.
This might occurs since the 2FGL catalog was not built using a blind
all-sky search \citep{nolan2012} and intrinsic source variability, or
artifacts due to the $\gamma$-ray background, could affect the above
estimates making them lower.

\section{Summary and Conclusions}
\label{sect:05}

We applied an association procedure based on the peculiar IR
properties of $\gamma$-ray emitting blazars to search for $\gamma$-ray
blazar candidates among the 382 unidentified BAT sources in the
3PBC.
Using this method, we obtained a preliminary list of \wse\ candidates
distributed between unclassified and unassociated 3PBC sources.
We searched in the literature the information about their
multifrequency emission properties and complemented it with the
analysis of \sw~observations, when available.
Consequently, we established more restrictive selection criteria
requiring the existence of consistent radio and soft X-ray emission
for the candidates selected by our association procedure; moreover, we
excluded candidates with an extended radio emission.

We obtained a list of 24 $\gamma$-ray blazar candidates: 5 of class~A
($\sim$21\%), 13 of class~B ($\sim$54\%), and 6 of class~C (25\%).
Only one of them corresponds to a 3PBC unassociated source.
Further three candidates are promising as a compact radio emission,
essential for a blazar classification according to the \bz criteria,
has been reported in the NVSS but they have never been observed in the
X rays.
We consider a very low probability of chance coincidence for our list
of candidates, of the order of 0.1\% for class~A candidates and of 1\%
for class~B and class~C candidates.

We note that the fraction of blazars included in the 3PBC and detected
by \frm-LAT is $\sim$~6\% of the identified 3PBC objects.
Assuming, in a first approximation, the same fraction of $\gamma$-ray
blazars among the unidentified 3PBC sources we would obtain 21
objects, in very good agreement with the number of candidates that we
selected.
At the same time, we also note that our results cannot rule out the
possibility that the hard X-ray emission from unidentified BAT sources
is due to a Seyfert galaxy or some other obscured AGN, rather than to
the blazar candidate provided by our method.

\vspace{0.5cm}

\acknowledgements

The authors are grateful to the referee for the helpful comments aimed
at improving this paper.
The work has been supported by the ASI grant I/011/07/0 and by the NASA
grant NNX12AO97G.
R. D'Abrusco gratefully acknowledges the financial support of the US
Virtual Astronomical Observatory, which is sponsored by the National
Science Foundation and the National Aeronautics and Space
Administration.
H.A. Smith acknowledges partial support from JPL/NASA contract 1455432-717437.
The work by G. Tosti is supported by the ASI/INAF contract I/005/12/0.
TOPCAT\footnote{\underline{http://www.star.bris.ac.uk/$\sim$mbt/topcat/}}
\cite{taylor2005} and SAOImage DS9 were used extensively in this
work for the preparation and manipulation of the tabular data and the
images.
This research has made use of data obtained from the High Energy
Astrophysics Science Archive Research Center (HEASARC) provided by
NASA's Goddard Space Flight Center; the SIMBAD database operated at
CDS, Strasbourg, France; the NASA/IPAC Extragalactic Database (NED)
operated by the Jet Propulsion Laboratory, California Institute of
Technology, under contract with the National Aeronautics and Space
Administration.
Part of this work is based on archival data, software or on-line
services provided by the ASI Science Data Center.
This publication makes use of data products from the Wide-field
Infrared Survey Explorer, which is a joint project of the University
of California, Los Angeles, and the Jet Propulsion
Laboratory/California Institute of Technology, funded by the National
Aeronautics and Space Administration.
6dF optical spectra are taken from the Final Release of 6dFGS
\cite{jones2004,jones2009}.  
Funding for SDSS-III has been provided by the Alfred P. Sloan
Foundation, the Participating Institutions, the National Science
Foundation, and the U.S. Department of Energy Office of Science. 
The SDSS-III web site is http://www.sdss3.org/.
SDSS-III is managed by the Astrophysical Research Consortium for the
Participating Institutions of the SDSS-III Collaboration including the
University of Arizona, the Brazilian Participation Group, Brookhaven
National Laboratory, University of Cambridge, Carnegie Mellon
University, University of Florida, the French Participation Group, the
German Participation Group, Harvard University, the Instituto de
Astrofisica de Canarias, the Michigan State/Notre Dame/JINA
Participation Group, Johns Hopkins University, Lawrence Berkeley
National Laboratory, Max Planck Institute for Astrophysics, Max Planck
Institute for Extraterrestrial Physics, New Mexico State University,
New York University, Ohio State University, Pennsylvania State
University, University of Portsmouth, Princeton University, the
Spanish Participation Group, University of Tokyo, University of Utah,
Vanderbilt University, University of Virginia, University of
Washington, and Yale University.

\appendix
\section{Some details about the association method}
\label{app:01}

We provide here the basic details of the association method adopted in
our analysis to search for $\gamma$-ray blazar candidates among the
unidentified BAT sources.
A complete description is given in Massaro et~al.~(2012a,b) and
additional details are also illustrated in D'Abrusco et~al.~(2013).

Our procedure is based on the peculiar infrared properties of the
$\gamma$-ray blazars that allow to distinguish them, at a high
confidence level, from other Galactic and extragalactic sources
dominated by thermal emission (see \citealp{dabrusco2012}).
The infrared counterparts, detected in all of the four \wse\ bands, of
the \frm-LAT blazars included in the 2LAC \cite{ackermann2011} locate
in a very narrow region of the IR color-color space that has been
named as the {\it \wse\ Gamma-ray Strip}.
We developed a parameterization of the strip that can be used to
search for $\gamma$-ray blazar candidates.
This parameterization is described in terms of the projections of the
strip in each of the three independent 2-dimensional planes of the IR
color-color space.
For each projection we defined the boundaries that include at least
90\% of the corresponding $\gamma$-ray blazars.
Then, we defined a parameter to characterize the position of a generic
IR source with respect to these boundaries.
For example, in the [3.4]-[4.6]-[12]~$\mu$m diagram we defined the
$s_{12}$ parameter: its value ranges in the 0--1 interval, with zero
corresponding to a source definitely outside these boundaries.  
In an analogous way we defined $s_{23}$ in the [4.6]-[12]-[22] $\mu$m
and $s_{34}$ in the [3.4]-[4.6]-[12]-[22] $\mu$m diagrams,
respectively.
Finally, these values are combined to define a single strip parameter
in the 3D \wse\ color-color space:
\begin{equation}
\label{spar}
s = (s_{12}\,s_{23}\,s_{34})^{1/3}.
\end{equation}
Thus, infrared sources that lie out of the boundaries of the
\wse\ Gamma-ray Strip have still $s$=0 as they do not have blazar-like
IR colors, while sources with $s$ close to 1 are likely $\gamma$-ray
blazar candidates.
By definition, $s$ is weighted for the errors on all the IR colors
and is normalized to range in the 0--1 interval.
Moreover, BL~Lacs and FSRQs cluster in distinct subregions of the IR
color-color space: for this reason it is possible to compute, for
each source, the $s_b$ parameter which is relevant to the part of the
strip including the majority of BL~Lacs, and $s_q$ relevant to the
region that mostly includes FSRQs (see also \citealp{massarof2012b}
for further details).

We selected random regions of the sky, at high Galactic latitude
with a total area of $\sim$~19 deg$^2$, and considered a sample of
10311 generic IR sources detected in all the four \wse\ bands like the
blazars in the Gamma-ray Strip.
We evaluated the $s_b$ and $s_q$ parameters for both the generic IR
sources and these $\gamma$-ray blazars, and computed the corresponding
distributions.
From their comparison we established the thresholds to define
three classes of $\gamma$-ray blazar candidates: 
\begin{center}
\begin{description} 
\item{class A:} $0.10 < s_b < 1.00$ AND $0.55 < s_q < 1.00$;
\item{class B:} $0.45 < s_b < 1.00$ OR \,\, $0.60 < s_q < 1.00$;
\item{class C}: $0.15 < s_b < 0.45$ OR $0.15 < s_q < 0.60$.
\end{description} 
\end{center}
All the \wse\ sources with $s_b < 0.15$ and $s_q < 0.15$ are
considered outliers of the \wse\ Gamma-ray Strip.
The above definition of the thresholds corresponds to an improvement
with respect to the previous one \cite{massarof2012b} as it is based
on the larger sample of \wse\ conterparts (610~vs~284) associated to
\frm-LAT detected blazars after the release of the \wse\ full
archive\footnote{http://wise2.ipac.caltech.edu/docs/release/allsky/}
available since March 2012 \citep[see also][]{cutri2012} and
containing the attributes for 563,921,584 point-like and resolved
objects.
The association method takes into account the correction for the
Galactic extinction for all the \wse\ magnitudes according to
Draine~(2003).
As shown in D'Abrusco et~al.~(2013) this affects only, and in a marginal
way, the [3.4]-[4.6] color index: in particular, for more than 95\%
of the blazars on the \wse\ Gamma-ray Strip the correction is less
than $\sim$3\%.

We remark that sources of class~A, with relatively higher values of
the $s_b$ and $s_q$ parameters, are in the innermost region of the
\wse\ Gamma-ray strip and in this sense can be considered more
reliable {\it candidates} with respect to sources of class~C
\cite{massarof2012b}.
However, this does not imply that sources of class~C are bad
candidates.
In fact, 559 out of 610 ($\sim$91.6\%) 2LAC blazars with a IR
counterpart detected in all of the four \wse\ bands have been
confirmed by our method.
Among them there are 134 sources of class~A ($\simeq$~24\%), 170
sources of class~B ($\simeq$~30\%), and 255 of class~C
($\simeq$~46\%).

On the basis of the described parameterization and the definition of
the threshold that establish three different classes for the
$\gamma$-ray blazar candidates, the method \cite{massarof2012a}
defines for each high-energy source a circular searching region
centered at its position with a radius equal to the positional
uncertainty at 99\% confidence level; then, all the \wse\ sources
within this searching region are ranked according to their $s$
parameter.
Of the obtained list of candidates we indicate, as a rule, our best
choice as the \wse\ source of higher class with the smallest angular
separation from the position of the high-energy source.

{}

\newpage

\begin{table}
\caption{The list of correspondences between unidentified 3PBC hard
  X-ray sources and \wse\ $\gamma$-ray blazar candidates. 1) The name
  of the source and 2) the counterpart provided in the 3PBC; 3) the
  \wse\ name, 4) the class of the candidate as designated by our
  association procedure and 5) its distance $d$ from the 3PBC
  counterpart; 6) the redshift $z$; 7) the radio flux density $F_r$;
  8) the $u-r$ color index as derived from SDSS-DR9; 9) the
  comparison between the positions of the \wse\ (small inner circle)
  and of the soft X-ray source. Further details are given in
  Section~\ref{sect:02}, in Section~\ref{sect:04:2} and in the
  Appendix.}
\label{table:2}
\begin{center}
\begin{tabular}{ccccccccc}
\hline
~\\
3PBC name      &   3PBC counterpart        &        \wse\ name      & class &        d       &    $z$   &  $F_r$   &  $u-r$  & IR/X \\ 
               &                           &                       &       &    (arcsec)    &          & (mJy)    &  (mag)  &      \\
~\\                                                                                                                 
\hline                                                                                                              
~\\                                                                                                                 
J0056.9$+$6401 & NVSS~J005712$+$635942     & J005712.84$+$635942.8 &   B   &   0.2          &  -       &   16.3~n &     -   &  \C  \\ 
J0109.3$+$0859 & 1RXS~J010929.7$+$08573    & J010930.21$+$085748.1 &   A   &  12.6          &  -       &   11.2~n &    1.99 &  \D  \\
J0137.4$+$5814 & RX~J0137.7$+$5814         & J013750.47$+$581411.3 &   C   &   0.2          &  -       &  170.7~n &    0.73 &  \C  \\ 
J0207.9$+$8410 & CRATES~J0207$+$8411       & J020713.45$+$841119.1 &   C   &   0.1          &  -       &  100.1~n &     -   &  \C  \\ 
J0312.0$+$5027 & IRAS~03084$+$5017         & J031202.97$+$502914.6 &   A   &   0.9          &  -       &   12.6~n &     -   &  \C  \\ 
J0356.2$-$6252 & 2MASX~J03561995$-$6251391 & J035619.96$-$625139.2 &   B   &   0.5          & 0.108~\1 &   20.4~s &     -   &  \C  \\ 
J0359.6$+$5059 & 4C~50.11                  & J035929.74$+$505750.1 &   B   &   0.1          & 1.52~\2  & 4296.6~n &    0.34 &  \C  \\ 
J0413.2$+$1659 & CRATES~J0413$+$1659       & J041322.32$+$165951.1 &   A   &   0.1          &  -       &   94.0~n & $-$0.03 &  \C  \\ 
J0421.1$+$2602 & 1RXS~J042054.9$+$26050    & J042056.00$+$260450.1 &   A   &  19.7          &  -       &    5.4~n &    1.91 &  \D  \\
J0440.8$+$2739 & 2MASX~J04404770$+$2739466 & J044047.72$+$273947.1 &   B   &   0.1          &  -       &    3.1~n &     -   &  \C  \\ 
J0459.9$+$2704 & 4C~27.14                  & J045956.08$+$270602.1 &   B   &   1.1          &  -       &  927.0~n &     -   &  \C  \\ 
J0612.2$-$4645 & 1RXS~J061227.3$-$464725   & J061226.91$-$464718.5 &   B   &   8.0          &  -       &  216.2~s &     -   &  \C  \\
J1240.1$+$3501 & 1RXS~J124020.3$+$350303   & J124021.14$+$350259.0 &   C   &   3.1          & 1.199~\3 &  230.8~n &    0.20 &  \D  \\
J1448.7$-$4007 & CRATES~J1448$-$4008       & J144850.99$-$400845.6 &   A   &  19.4          &  -       &   49.0~n &     -   &  \C  \\ 
J1508.7$-$4952 & 1RXS~J150839.0$-$495304   & J150838.93$-$495302.2 &   C   &   2.4          &  -       &  595.0~p &     -   &  \C  \\ 
J1540.1$+$1414 & CRATES~J1540$+$1411       & J154007.85$+$141137.2 &   B   &  30.6          & 0.119~\3 &   21.5~f &    0.35 &  \C  \\ 
J1742.0$-$6053 & 1RXS~J174201.5$-$605514   & J174201.50$-$605512.1 &   B   &   1.8          & 0.41~\4  &   97.7~s &     -   &  \C  \\
J1745.6$+$2907 & SWIFT~J1745.4$+$2906      & J174538.26$+$290822.2 &   B   &   0.2          & 0.111~\5 &   13.3~n &     -   &  \C  \\ 
J1812.8$-$0648 & PMN~J1812$-$0648          & J181250.94$-$064825.4 &   B   &   1.6          &  -       & 1342.2~n &     -   &  \C  \\ 
J1926.5$+$4132 & 1RXS~J192630.6$+$413314   & J192630.21$+$413305.0 &   B   &  10.4          &  -       &    2.8~n &     -   &  \C  \\ 
J2010.3$-$2524 & 1RXS~J201020.0$-$252356   & J201019.76$-$252359.1 &   C   &   4.2          &  -       &   58.9~n &     -   &  \C  \\ 
J2030.1$+$7609 &                           & J202952.72$+$761139.2 &   C   & 159.7$^{\star}$ &  -       &  100.0~n &    0.36 &  \C  \\ 
J2117.7$+$5138 & IGR~J21178$+$5139         & J211747.70$+$513856.8 &   B   &   4.2          &  -       &    9.1~n &     -   &  \C  \\
J2214.0$-$2557 & 2MASX~J22140917$-$2557487 & J221409.17$-$255749.1 &   B   &   0.2          & 0.052~\1 &    3.4~n &     -   &  \C  \\ 
~\\
\hline
~\\
\end{tabular}
\end{center}
\scriptsize{Redshift: 1) 6dF \cite{jones2009}; 2) Agudo et~al.~(2007);
  3) SDSS-DR9 \cite{paris2012}; 4) Burgess et~al.~(2006); 5) Tueller
  et~al.~(2008). \\ Radio flux density: f) \cite{white1997}; n) NVSS
  \cite{condon1998}; p) PMN \cite{wright1994}; s) SUMMS
  \cite{mauch2003}.\\ ($\star$) this value represents the distance of
  the \wse\ source from the 3PBC source.}
\end{table}

\begin{table}
\caption{Optical-UV photometry of \wse\ $\gamma$-ray blazar candidates
  observed by \sw-UVOT. 1) The \wse\ name of the candidate, 2) the
  Galactic latitude $b$, 3) the E(B-V) value with 4) its error, 5) the
  identifier of the \sw~observation, 6) the date of the observation,
  7) the UVOT filter, 8) the exposure, 9) the magnitude with 10) the
  corresponding error. The values without errors have to be considered
  as upper limits.}
\label{table:3}
\begin{center}
\begin{tabular}{cccccccccc}
\hline
~\\
     \wse\ name        &    b     & E(B-V) & error  &    obsID    &     date    & filter & exposure & magnitude & error \\
                      &  (deg)   & (mag)  & (mag)  &             &             &        &   (s)    &   (mag)   & (mag) \\
~\\                                          
\hline                                                                                                              
~\\                                          
J005712.84$+$635942.8 &  $+$1.13 &  1.46  &  0.03  & 00041144001 & 2010 Aug 30 &   M2   &   6562   &   21.6    &       \\
         $''$         &   $''$   &  $''$  &  $''$  &     $''$    & 2010 Aug 29 &   W2   &   3194   &   21.1    &  0.3  \\
         $''$         &   $''$   &  $''$  &  $''$  & 00041144002 & 2010 Aug 31 &   W1   &   2698   &   20.5    &  0.2  \\
J013750.47$+$581411.3 &  $-$4.09 &  0.53  &  0.01  & 00041273001 & 2010 Sep 04 &   W1   &   1163   &   20.1    &  0.3  \\
         $''$         &   $''$   &  $''$  &  $''$  & 00041273002 & 2010 Oct 22 &   W1   &   3358   &   19.8    &  0.1  \\  
J020713.45$+$841119.1 & $+$21.62 &  0.11  &  0.01  & 00039241001 & 2009 Sep 12 &   M2   &   5345   &   17.7    &  0.1  \\
J031202.97$+$502914.6 &  $-$6.38 &  0.71  &  0.01  & 00038026001 & 2008 Oct 11 &   M2   &   1557   &   20.3    &  0.3  \\
         $''$         &   $''$   &  $''$  &  $''$  & 00038026002 & 2008 Dec 13 &   W1   &   6719   &   18.9    &  0.1  \\
J035619.96$-$625139.2 & $-$43.37 &  0.05  &  0.01  & 00037304002 & 2008 Oct 10 &   M2   &   5044   &   18.5    &  0.1  \\
J035929.74$+$505750.1 &  $-$1.60 &  1.51  &  0.05  & 00030879001 & 2007 Jan 25 &    U   &   3449   &   21.5    &       \\
         $''$         &   $''$   &  $''$  &  $''$  & 00030879002 & 2007 Jan 30 &    U   &   1519   &   21.0    &       \\
         $''$         &   $''$   &  $''$  &  $''$  & 00030879003 & 2007 Feb 07 &    U   &   3776   &   20.9    &  0.2  \\
         $''$         &   $''$   &  $''$  &  $''$  & 00030879004 & 2007 Feb 13 &    U   &   5852   &   21.3    &  0.3  \\
         $''$         &   $''$   &  $''$  &  $''$  & 00030879005 & 2007 Feb 15 &    U   &   5189   &   21.4    &       \\
         $''$         &   $''$   &  $''$  &  $''$  & 00036308001 & 2007 Nov 05 &   W2   &   8625   &   22.2    &       \\
         $''$         &   $''$   &  $''$  &  $''$  & 00036308002 & 2007 Dec 02 &    U   &   8175   &   22.0    &       \\
         $''$         &   $''$   &  $''$  &  $''$  &     $''$    &   $''$      &   W2   &   3463   &   21.7    &       \\
         $''$         &   $''$   &  $''$  &  $''$  & 00036308003 & 2008 Aug 13 &   W1   &   2526   &   21.0    &       \\
         $''$         &   $''$   &  $''$  &  $''$  & 00036308004 & 2008 Oct 23 &   M2   &   2145   &   21.0    &       \\
         $''$         &   $''$   &  $''$  &  $''$  & 00036308005 & 2008 Oct 24 &    U   &   5725   &   21.5    &       \\
         $''$         &   $''$   &  $''$  &  $''$  &     $''$    &   $''$      &   W1   &   3419   &   20.6    &  0.2  \\
         $''$         &   $''$   &  $''$  &  $''$  &     $''$    &   $''$      &   W2   &    208   &   19.9    &       \\
         $''$         &   $''$   &  $''$  &  $''$  & 00036308006 & 2008 Oct 27 &   W1   &   1034   &   20.7    &       \\
         $''$         &   $''$   &  $''$  &  $''$  & 00030879006 & 2010 Apr 10 &    U   &   3716   &   21.2    &       \\
         $''$         &   $''$   &  $''$  &  $''$  & 00030879007 & 2010 Nov 01 &    V   &    320   &   19.6    &       \\
         $''$         &   $''$   &  $''$  &  $''$  & 00030879007 & 2010 Nov 01 &    B   &    320   &   20.5    &       \\
         $''$         &   $''$   &  $''$  &  $''$  & 00030879007 & 2010 Nov 01 &    U   &    320   &   20.2    &       \\
         $''$         &   $''$   &  $''$  &  $''$  & 00030879007 & 2010 Nov 01 &   W1   &    641   &   20.5    &       \\
         $''$         &   $''$   &  $''$  &  $''$  & 00030879007 & 2010 Nov 01 &   M2   &    908   &   20.6    &       \\
         $''$         &   $''$   &  $''$  &  $''$  & 00030879007 & 2010 Nov 01 &   W2   &   1283   &   21.1    &  0.4  \\
J041322.32$+$165951.1 & $-$24.10 &  0.66  &  0.03  & 00090151002 & 2009 Aug 02 &   W2   &   5717   &   19.3    &  0.1  \\
J044047.72$+$273947.1 & $-$12.34 &  0.73  &  0.02  & 00040910001 & 2010 Jul 22 &   W1   &   5836   &   18.9    &  0.1  \\
         $''$         &   $''$   &  $''$  &  $''$  & 00040910002 & 2010 Jul 23 &    U   &   4253   &   17.7    &  0.1  \\
         $''$         &   $''$   &  $''$  &  $''$  & 00040694001 & 2010 Aug 07 &   W1   &   1029   &   19.0    &  0.1  \\
         $''$         &   $''$   &  $''$  &  $''$  & 00040910003 & 2010 Aug 07 &   W1   &   1533   &   18.9    &  0.1  \\
         $''$         &   $''$   &  $''$  &  $''$  & 00040694002 & 2011 Jan 09 &   M2   &    878   &   20.6    &  0.4  \\
         $''$         &   $''$   &  $''$  &  $''$  & 00040694003 & 2011 Jan 14 &   W1   &   1694   &   19.0    &  0.1  \\
         $''$         &   $''$   &  $''$  &  $''$  & 00040910004 & 2012 Jul 22 &   W1   &   2518   &   19.0    &  0.1  \\
         $''$         &   $''$   &  $''$  &  $''$  &     $''$    & 2012 Jul 22 &   M2   &   9756   &   21.1    &  0.2  \\
         $''$         &   $''$   &  $''$  &  $''$  & 00040910005 & 2012 Jul 24 &    U   &  19133   &   17.8    &  0.1  \\
         $''$         &   $''$   &  $''$  &  $''$  & 00040910006 & 2012 Jul 25 &   W2   &  12095   &   20.7    &  0.1  \\
J045956.08$+$270602.1 &  $-$9.36 &  1.13  &  0.04  & 00038030001 & 2008 Nov 18 &    U   &    477   &   20.3    &  0.4  \\    
         $''$         &   $''$   &  $''$  &  $''$  & 00038030002 & 2009 Jan 13 &    U   &   4292   &   20.4    &  0.2  \\
         $''$         &   $''$   &  $''$  &  $''$  & 00038030003 & 2009 Mar 10 &    U   &   1444   &   20.7    &       \\
         $''$         &   $''$   &  $''$  &  $''$  & 00038030004 & 2009 Aug 03 &   M2   &   4927   &   21.5    &       \\
J061226.91$-$464718.5 & $-$25.82 &  0.06  &  0.01  & 00041159001 & 2010 Jul 06 &   W1   &   1007   &   18.9    &  0.1  \\ 
         $''$         &   $''$   &  $''$  &  $''$  & 00041159002 & 2010 Nov 22 &   M2   &   2297   &   18.8    &  0.1  \\
         $''$         &   $''$   &  $''$  &  $''$  & 00041159003 & 2010 Dec 04 &   M2   &   6668   &   18.8    &  0.1  \\   
~\\
\hline
\end{tabular}
\end{center}
\end{table}

\setcounter{table}{2}                                                         
\begin{table}
\caption{continued.}
\begin{center}
\begin{tabular}{cccccccccc}
\hline
~\\
     \wse\ name       &    b     & E(B-V) & error  &    obsID    &     date    & filter & exposure & magnitude & error \\
                      &  (deg)   & (mag)  & (mag)  &             &             &        &   (s)    &   (mag)   & (mag) \\
~\\                                              
\hline                                                                                                                 
~\\                                             
J144850.99$-$400845.6 & $+$17.40 &  0.11  &  0.01  & 00045398001 & 2011 Mar 19 &   W1   &    790   &   16.6    &  0.1  \\
         $''$         &   $''$   &  $''$  &  $''$  & 00045398002 & 2011 Mar 21 &   W2   &   2119   &   17.2    &  0.1  \\
         $''$         &   $''$   &  $''$  &  $''$  & 00045398003 & 2011 Mar 25 &   W2   &   5822   &   17.1    &  0.1  \\
J150838.93$-$495302.2 &  $+$7.18 &  0.38  &  0.01  & 00037996001 & 2010 Jun 21 &    U   &   2312   &   18.8    &  0.1  \\
         $''$         &   $''$   &  $''$  &  $''$  &     $''$    & 2010 Jun 20 &   W1   &    212   &   19.6    &  0.4  \\
         $''$         &   $''$   &  $''$  &  $''$  & 00037996002 & 2010 Jun 23 &   M2   &   7568   &   20.5    &  0.1  \\
         $''$         &   $''$   &  $''$  &  $''$  & 00041522001 & 2012 May 05 &    U   &    351   &   18.4    &  0.1  \\
         $''$         &   $''$   &  $''$  &  $''$  & 00041522002 & 2012 May 07 &   M2   &    367   &   19.9    &       \\
J154007.85$+$141137.2 & $+$48.70 &  0.05  &  0.01  & 00039843001 & 2009 Oct 01 &   W2   &   6723   &   15.8    &  0.1  \\
         $''$         &   $''$   &  $''$  &  $''$  & 00039843003 & 2010 Jul 02 &   W1   &   3763   &   15.6    &  0.1  \\
J174201.50$-$605512.1 & $-$15.63 &  0.08  &  0.01  & 00047123001 & 2011 Nov 10 &   W1   &    823   &   15.5    &  0.1  \\
         $''$         &   $''$   &  $''$  &  $''$  & 00047123002 & 2011 Nov 11 &   W2   &   3301   &   15.4    &  0.1  \\
         $''$         &   $''$   &  $''$  &  $''$  &     $''$    &   $''$      &    U   &   1813   &   15.4    &  0.1  \\
         $''$         &   $''$   &  $''$  &  $''$  & 00047123003 & 2011 Nov 14 &   W1   &   1105   &   15.6    &  0.1  \\
         $''$         &   $''$   &  $''$  &  $''$  & 00047123004 & 2011 Nov 17 &   M2   &   2782   &   15.5    &  0.1  \\
J174538.26$+$290822.2 & $+$26.24 &  0.05  &  0.01  & 00035273001 & 2005 Dec 11 &    V   &    989   &   15.7    &  0.1  \\
         $''$         &   $''$   &  $''$  &  $''$  &     $''$    &   $''$      &    B   &    423   &   16.5    &  0.1  \\
         $''$         &   $''$   &  $''$  &  $''$  &     $''$    &   $''$      &    U   &    573   &   15.4    &  0.1  \\
         $''$         &   $''$   &  $''$  &  $''$  &     $''$    &   $''$      &   W1   &   1431   &   15.5    &  0.1  \\
         $''$         &   $''$   &  $''$  &  $''$  &     $''$    &   $''$      &   M2   &   2963   &   15.6    &  0.1  \\
         $''$         &   $''$   &  $''$  &  $''$  &     $''$    &   $''$      &   W2   &   3959   &   15.4    &  0.1  \\
J181250.94$-$064825.4 &  $+$5.40 &  1.34  &  0.02  & 00045404001 & 2012 Feb 11 &    U   &   1291   &   20.5    &       \\
         $''$         &   $''$   &  $''$  &  $''$  & 00045404002 & 2012 Nov 05 &    U   &   1817   &   20.8    &       \\
         $''$         &   $''$   &  $''$  &  $''$  & 00045404003 & 2012 Nov 06 &   W2   &    497   &   20.4    &       \\
J192630.21$+$413305.0 & $+$11.59 &  0.10  &  0.01  & 00039097001 & 2009 Jun 12 &   M2   &   3112   &   18.3    &  0.1  \\
         $''$         &   $''$   &  $''$  &  $''$  & 00039097003 & 2009 Jul 09 &   W2   &   3016   &   17.6    &  0.1  \\
         $''$         &   $''$   &  $''$  &  $''$  & 00039097006 & 2009 Aug 25 &    U   &   9143   &   16.9    &  0.1  \\
         $''$         &   $''$   &  $''$  &  $''$  & 00039097007 & 2009 Aug 26 &   W2   &  12137   &   17.4    &  0.1  \\
J201019.76$-$252359.1 & $-$27.83 &  0.16  &  0.01  & 00090901001 & 2010 Jun 26 &   W2   &    797   &   19.5    &  0.1  \\
         $''$         &   $''$   &  $''$  &  $''$  & 00040717001 & 2010 Nov 19 &   W1   &   2200   &   19.1    &  0.1  \\
         $''$         &   $''$   &  $''$  &  $''$  & 00041102001 & 2010 Nov 24 &    U   &   3170   &   18.8    &  0.1  \\
         $''$         &   $''$   &  $''$  &  $''$  &     $''$    & 2010 Nov 25 &   W2   &   5134   &   19.4    &  0.1  \\
         $''$         &   $''$   &  $''$  &  $''$  & 00040717003 & 2012 Aug 24 &   W1   &    887   &   19.3    &  0.1  \\
J202952.72$+$761139.2 & $+$20.75 &  0.25  &  0.01  & 00046326001 & 2012 May 30 &   W2   &   1128   &   21.0    &       \\
         $''$         &   $''$   &  $''$  &  $''$  & 00046326002 & 2012 Jun 03 &   W2   &    431   &   20.5    &       \\
         $''$         &   $''$   &  $''$  &  $''$  & 00046326003 & 2012 Jun 05 &   W1   &    287   &   20.0    &       \\
         $''$         &   $''$   &  $''$  &  $''$  & 00046326004 & 2012 Jun 10 &    U   &   1333   &   20.4    &  0.3  \\
         $''$         &   $''$   &  $''$  &  $''$  &     $''$    &   $''$      &   W2   &    435   &   20.4    &       \\
         $''$         &   $''$   &  $''$  &  $''$  & 00046326005 & 2012 Jun 11 &   M2   &    998   &   20.8    &       \\
J211747.70$+$513856.8 &  $+$1.63 &  3.12  &  0.21  & 00037140001 & 2007 May 20 &    U   &   7076   &   21.4    &       \\ 
         $''$         &   $''$   &  $''$  &  $''$  & 00037140002 & 2007 May 23 &   W1   &   5857   &   21.3    &  0.3  \\
J221409.17$-$255749.1 & $-$54.95 &  0.02  &  0.01  & 00041121001 & 2011 Apr 03 &   M2   &   1013   &   17.2    &  0.1  \\
         $''$         &   $''$   &  $''$  &  $''$  & 00041121002 & 2011 Apr 05 &    U   &   3867   &   16.8    &  0.1  \\
         $''$         &   $''$   &  $''$  &  $''$  & 00041121003 & 2011 May 09 &   M2   &    734   &   17.2    &  0.1  \\
         $''$         &   $''$   &  $''$  &  $''$  &     $''$    & 2011 May 08 &   W2   &    954   &   17.4    &  0.1  \\
         $''$         &   $''$   &  $''$  &  $''$  & 00041121004 & 2011 Jun 23 &   W1   &    740   &   17.2    &  0.1  \\
         $''$         &   $''$   &  $''$  &  $''$  & 00041121005 & 2011 Jun 25 &   W2   &    412   &   17.3    &  0.1  \\
         $''$         &   $''$   &  $''$  &  $''$  & 00041121006 & 2011 Jun 29 &   W2   &   2681   &   17.3    &  0.1  \\
~\\
\hline
\end{tabular}
\end{center}
\end{table}

\end{document}